# Astro2020 Science White Paper

# Engaging Citizen Scientists to Keep Transit Times Fresh and Ensure the Efficient Use of Transiting Exoplanet Characterization Missions

**Thematic Areas:** ☒ Planetary Systems

**Principal Author:**
Name: Robert T. Zellem
Institution: Jet Propulsion Laboratory, California Institute of Technology
Email: rzellem@jpl.nasa.gov
Phone: 626-379-9418

**Co-authors:**
Anya Biferno, Jet Propulsion Laboratory, California Institute of Technology Anya.A.Biferno@jpl.nasa.gov, David R. Ciardi, NASA Exoplanet Science Institute/California Institute of Technology, ciardi@ipac.caltech.edu, Mary Dussault, Center for Astrophysics | Harvard & Smithsonian, mdussault@cfa.harvard.edu, Laura Peticolas, Sonoma State University, laurap@universe.sonoma.edu, Martin Fowler, Citizen Scientist, danebury216@hotmail.co.uk, Kyle A. Pearson, Lunar and Planetary Laboratory, University of Arizona, pearsonk@lpl.arizona.edu, Wilfred Gee, Macquarie University, wilfred.gee@students.mq.edu.au, Rachel Zimmerman-Brachman, Jet Propulsion Laboratory, California Institute of Technology, Rachel.Zimmerman-Brachman@jpl.nasa.gov, Denise Smith, Space Telescope Science Institute, dsmith@stsci.edu, Lynn Cominsky, Sonoma State University, lynnc@universe.sonoma.edu, Gael M. Roudier, Jet Propulsion Laboratory, California Institute of Technology, gael.m.roudier@jpl.nasa.gov, Brandon Lawton, Space Telesope Science Institute, lawton@stsci.edu, Robert Baer, Southern Illinois University Carbondale, Citizen CATE Experiment, rbaer@siu.edu, Diana Dragomir, Massachusetts Institute of Technology, NASA Hubble Fellow, dragomir@space.mit.edu, Nemanja Jovanovic, California Institute of Technology, nem@caltech.edu, Marc Kuchner, NASA Goddard Space Flight Center, Marc.Kuchner@nasa.gov, Frank Sienkiewicz, Center for Astrophysics | Harvard & Smithsonian, fsienkiewicz@cfa.harvard.edu, and Josh Walawender, W. M. Keck Observatory, jwalawender@keck.hawaii.edu

## Introduction

TESS is predicted to discover 10,000+ transiting exoplanets (Barclay et al. 2018), providing thousands of bright targets with large transit depths ideal for in-depth follow-up spectroscopic characterization by Hubble, JWST, and other next-generation platforms (Zellem et al. 2017). However, TESS will sample some of its targets for only 27 days and, even in the continuous viewing zone, TESS's nominal mission will end before the launch of JWST and ARIEL (2021 and 2028, respectively). As a result, the mid-transit times of many of these targets can become "stale", i.e., the time of the next transit can become uncertain to the level where a significant amount of observing time must be invested to recover it. For example, a planet with an uncertainty of just 1 minute in both its orbital period and mid-transit time will have an uncertainty of ~15 hours in its mid-transit time in 10 years. Combating this problem requires follow-up from either ground- or space-based platforms (e.g., with Spitzer; Benneke et al. 2017) to reduce the uncertainties on a planet's period and mid-transit time. However, given that JWST will study tens to 100-200 transiting exoplanets in detail (Cowan et al. 2015; Kempton et al. 2018) and that ARIEL will survey ~1000 transiting exoplanets, a centralized effort is required for ephemerides maintenance.

This white paper advocates for the creation of a community-wide program to maintain precise mid-transit times of exoplanets that would likely be targeted by future platforms. Given the sheer number of targets that will require careful monitoring between now and the launch of the next generation of exoplanet characterization missions, this network will initially be devised as a citizen science project—focused on the numerous amateur astronomers, small universities and community colleges and high schools that have access to modest sized telescopes and off-the-shelf CCDs.

## Transit Timing Maintenance Is Critical for Characterization Surveys

If an exoplanet has a sufficiently large uncertainty in its mid-transit time or orbit period, then it will have a large uncertainty in its next expected mid-transit or mid-eclipse time. The next mid-transit or mid-eclipse time can be calculated via:

$$t_{next} = n_{orbit} \cdot P + T_{mid}$$

where $t_{next}$ is the time of the next transit, $P$ is the orbital period of the planet, $T_{mid}$ is the planet's mid-transit ephemeris, and $n_{orbit}$ is the number of orbits that have occurred between $t_{next}$ and $T_{mid}$. Performing error propagation of this equation, we find:

$$\text{var}(t_{next}) = n_{orbit}^2 \cdot \text{var}(P) + 2n_{orbit} \cdot \text{cov}(P, T_{mid}) + \text{var}(T_{mid})$$

Thus, due to the $n_{orbit}$ term, the uncertainty in measuring the next transit is most highly dependent upon the uncertainty in the orbital period $P$.

The current ~400 TESS Objects of Interest[1] (Fig. 1) have a median orbital period uncertainty of 0.51 minutes and a median mid-transit time uncertainty of 2.35 minutes. Assuming that these values are typical for TESS-observed transiting planets, then a representative TESS planet could have an uncertainty of 47.34 minutes after 1 year and 3.94 hours after 5 years (Fig. 2). Please note that these uncertainties are lower limits because it is assumed that the period and mid-transit times are

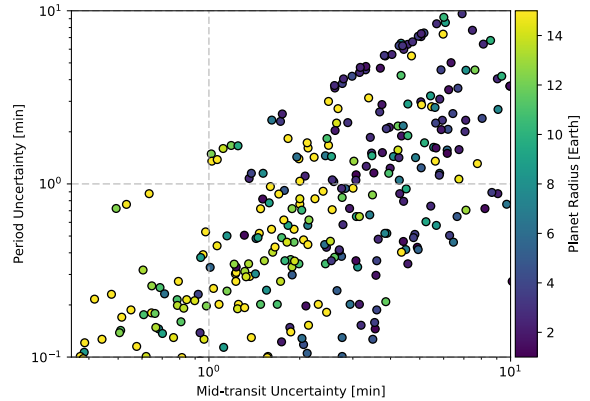

*Figure 1: Measured period uncertainties as a function of mid-transit uncertainties and planet radius of the ~400 planets currently in the TESS Objects of Interest catalog.*

---
1 https://exofop.ipac.caltech.edu/tess/view_toi.php

un-correlated–however, in reality that assumption is likely not true.

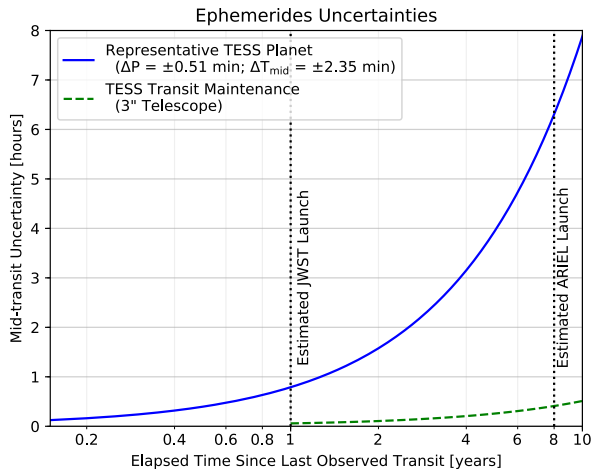

Figure 2: *The mid-transit uncertainty as a function of the elapsed time since last observed of a representative TESS planet (blue solid line). We assume that JWST launches ~1 year and ARIEL launches ~8 years after the end of TESS's primary mission operations (at x = 0 years). A single 3-inch telescope (scaled from measurements by a 61-inch telescope; Zellem et al. 2015) can greatly reduce the mid-transit uncertainty on a 9.34 H-mag planet (roughly the mean brightness of a typical transit target; Zellem et al. 2017), as denoted in dashed green.*

Thus, while TESS will measure highly-precise mid-transit times, if a target is observed only a handful of times in a particular TESS field (e.g., a planet with a long orbital period in a short-observation TESS field), then the planet will still have large uncertainties, particularly after many orbits. Therefore, orbital period and mid-transit maintenance through regular monitoring is necessary to keep these uncertainties small so they can be efficiently followed-up by other platforms, such as JWST.

**A Community of Citizen Scientists is Needed**

A community-wide citizen science effort to monitor transiting exoplanets with ground-based amateur telescopes would support these missions by alleviating observing overhead; here we advocate for its formation as the Exoplanet Timing Survey (ETS). Citizen scientists, in particular, provide a unique opportunity to the professional astronomical community: a large number of observers who are eager to aid NASA's mission goals and contribute to the observational needs of the professional community (e.g., Croll et al. 2011; Wiggins & Crowston 2011; Catlin-Groves 2012; Croll 2012; Franzoni & Sauermann 2014; Marshall et al. 2015; Kuchner et al. 2016; Watson et al. 2016). However just a single transit observation of a 9.34 H-mag TESS planet by a 3-inch diameter telescope can greatly reduce its 1-year-old mid-transit time uncertainty from 47.34 minutes to just 3.49 minutes (Fig. 2). This reduction in uncertainty is sufficient to keep this target's timing fresh ($\Delta t_{next} \leq 15$ mins) for the next 3.8 years.

A network of amateur astronomers could rapidly respond to new discoveries and high priority bright targets with large transit depths and monitor them, allowing larger telescopes to spend their time on other targets, like Earth-sized planets transiting dim M-dwarf stars. In addition, such a network could continue to monitor TESS fields, particularly those which will be observed by TESS for only ~27 days; thus, ETS has the potential to confirm planets with long orbital periods (P>27 days), which have only one TESS-observed transit.

Such a citizen science effort would also have the added benefit of increasing the scientific literacy of the community, increasing the number of highly-educated volunteers who can benefit NASA via their expertise, leveraging existing communities of citizen scientists or other enthusiasts, and connecting citizen scientists with NASA Subject Matter Experts. ETS would provide an authentic science opportunity as participants would collect observational data that will be provided to the science community for use in future observations. This facet is inspired by the Czech Astronomical Society's Transiting Exoplanets and Candidates[2] program and Exoplanet Transit Database[3].

The needed transit timing effort advocated here is complementary to the current on-going TESS Follow-Up Observation Program[4] (TFOP) which is eliminating TESS candidate false positives and confirming true TESS planetary systems. The TFOP utilizes ground-based time series photometry to eliminate blended eclipsing binaries, spectroscopy to obtain stellar parameters, high resolution imaging to identify potential binary systems, and precision radial velocities to determine the masses of the planets. The TFOP effort will help the community identify some of the best TESS planets for atmospheric characterization from the ground or with space facilities like JWST and ARIEL.

---

2 http://var2.astro.cz/EN/tresca/index.php
3 http://var2.astro.cz/ETD/
4 https://heasarc.gsfc.nasa.gov/docs/tess/followup.html

However, the TFOP effort is geared toward the mass determination of the planets. Without continued follow-up beyond the TESS mission, the ephemerides of the TESS planets will continue to grow stale. The work advocated here builds on the work of the TFOP – so that the transit midpoint times will be known sufficiently for future observations. The TFOP is necessary to identify the true planetary systems detected by TESS while ETS would maintain the quality of the ephemerides to ensure the suitability and efficiency of future characterization observations with JWST and ARIEL.

### Aspects of a Community Exoplanet Transit Timing Survey

ETS features four major facets: target identification, target observation and reduction, target global analysis, and publication of new ephemerides to the world community. Target identification, global analysis, and making the ephemerides public are fully-automated processes, designed to limit human-in-the-loop involvement, while target observation and reduction occurs entirely on the user-end, designed to enhance the users' experience in the ETS program.

#### Target Identification

Target selection would take advantage of the work already done by the professional community with automatic daily downloads from the NASA Exoplanet Archive for parameters on confirmed transiting planets and from the NExScI ExoFOP-TESS site for parameters on candidate planets identified by both the community and the TESS project. Each list would be further culled to select the targets with the largest uncertainties in terms of their mid-transit times and orbital periods. This list will then be ranked according to the transit figure of merit independently derived by Cowan et al. (2015), Zellem et al. (2017), Goyal et al. (2018), Kempton et al. (2018), and Morgan et al. (2018). This figure of merit weights targets preferentially with larger scale heights and brighter stars. Thus, it places higher priority on targets that are likely to be observed with HST, JWST, and other current and future telescopes for atmospheric characterization via transit spectroscopy; these same targets are also ideal for follow-up and routine monitoring with smaller, ground-based telescopes. Targets can also be overridden in the case of promoting a high-priority target that would not rank high (e.g., a super-earth with a large transit depth but a small scale height around a bright star) at the input of the exoplanet community or a mission. These ranked lists, and their associated transit times, would then be automatically made public daily and potentially also incorporated into a website available to citizen scientists (e.g., SciStarter).

#### Target Observation and Reduction

Using instructions and videos developed by ETS, each user would observe one of the identified targets. At the conclusion of the night, each observer would also reduce their own dataset for two reasons: first, it offloads computational requirements on the pipeline side of ETS by not requiring an official pipeline reduction of the data. Second, it would increase a user's educational experience and provide measurable learning outcomes by giving each user a full experience of taking and reducing their own data, thus giving them an authentic scientific experience and participating in the scientific process.

#### Target Global Analysis

Each citizen scientist would upload their reduced data, their measured transit depth, measured mid-transit time, and a stacked image of the field to the ETS website. With the exception of a single image of the field, no raw data will be uploaded to ETS to lessen the storage requirements on the system. A single image of the field would allow ETS to provide comparatively higher spatial resolution images (TESS has a 21" pixel scale while ground-based seeing-limited observations are typically on the order of 1-4") to the exoplanet community to search for blended pairs.

While a single user could potentially upload data of poor quality, crowdsourcing enables us to combine all observations and determine statistically weighted parameter solutions. ETS leverages itself via strength-in-numbers: the multitude of good observations will outnumber the bad ones. In order to maintain and increase participation, in addition to the daily tweets, each year articles would be written for amateur astronomy magazines, press releases would be developed during noteworthy targets, and local press would be contacted to generate news articles about the project when team members are in various towns for conferences or work meetings.

Whenever a new target or new data for a target is uploaded, the global means and uncertainties are automatically recalculated. The ETS pipeline will also calculate each planet's orbital period, enabling the search for transit timing variations and the discovery of new exoplanets (Holman & Murray 2005; Agol et al. 2005). Each planet's orbital period, transit depth, and mid-transit time will be published immediately on the ETS website, accessible by anyone: the amateur community, the professional exoplanet community, and the exoplanet archives, such as the NASA Exoplanet Archive and ExoFOP-TESS.

In addition to enabling the efficient use of NASA's near-future and future great observatories by leveraging recent work by Kepler, K2, and TESS, ETS would result in a multitude of published, peer-reviewed articles expanding beyond pure exoplanet science. Citizen scientists who contribute key data or analysis, would be included on published papers. Data gathered by citizen scientists and shared back with the public would follow the current standards on referencing such datasets.

*Recruiting Participants and Keeping Them Engaged*

As a citizen science effort, ETS represents a strong partnership between the science community and education and public engagement professionals. A comprehensive program to engage citizen scientists, and increase their numbers, will be designed leveraging community best practices and research conducted on creating strong, sustained engagement programs with the public. While initially aimed at the amateur astronomer and small colleges and universities, anyone can participate in ETS, even professional astronomers. Well-established citizen science portals, such as SciStarter, could potentially serve as pathways to participation in ETS. Platforms, such as Zooniverse or Kaggle, for analyzing the observations by citizen scientists is also a promising possibility (Watson et al. 2016).

ETS will also explore various methods to keep users engaged and actively participating. While individuals often participate in citizen science projects in order to contribute to science (Franzoni & Sauermann 2014; Marshall et al. 2015), ETS would potentially employ internet forums to foster community collaboration and a leaderboard to encourage friendly competition (e.g., Wald et al. 2016).

## NASA's Universe of Learning

The NASA's Universe of Learning (NASA's UoL[5]) program is well poised to develop ETS and ensure its success. NASA's UoL creates and delivers science- and audience-driven resources and experiences designed to engage learners of all ages and backgrounds in exploring the universe for themselves. Thus, NASA's UoL directly addresses NASA's Science Mission Directorate Citizen Science Policy[6]. Some of the elements of NASA's UoL that are particularly well suited to meet the needs of ETS are:

- MicroObservatory (microobservatory.org) is a network of 6-inch automated telescopes developed by the Center for Astrophysics | Harvard & Smithsonian that can be accessed at no charge

---

5 universe-of-learning.org; NASA's UoL is a partnership between the Space Telescope Science Institute, Caltech/IPAC, Jet Propulsion Laboratory, Smithsonian Astrophysical Observatory, and Sonoma State University, funded within the NASA Science Mission Directorate Science Activation (Sci-Act) program.
6 https://smd-prod.s3.amazonaws.com/science-red/s3fs-public/atoms/files/SPD-33%20Signed.pdf

over the Internet. All aspects of the telescopes' optical system, electronics, hardware, software, and graphic user interfaces have been optimized to enable educational use by novice observers, while preserving the powerful capabilities of remotely operated CCD imaging (Fig. 3).

- The Global Telescope Network (GTN; gtn.sonoma.edu) is a program developed and maintained by scientists, engineers, and educators at Sonoma State University to advance STEM learning and literacy by using internet-controllable telescopes to engage learners in authentic research experiences. GTN features the Gamma-ray Optical Robotic Telescope, a 14-inch robotic telescope in Northern California.

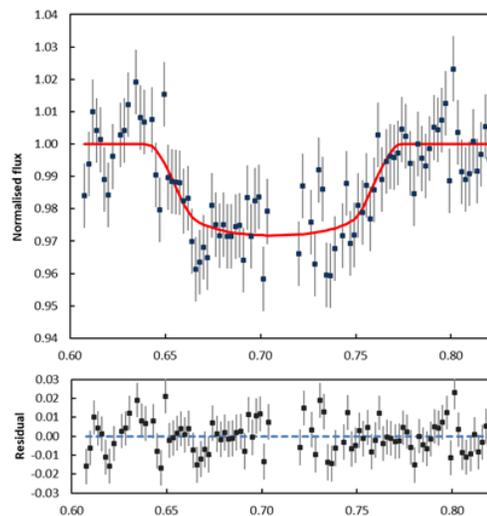

- The Panoptic Astronomical Networked Observatories for a Public Transiting Exoplanets Survey (PANOPTES; projectpanoptes.org) is a citizen science project which aims to build low cost, robotic telescopes which can be used to detect and follow-up transiting exoplanets. PANOPTES will establish a world-wide network of small cameras to monitor a large fraction of the sky to detect exoplanet transits and enable other discoveries.

*Figure 3: An example lightcurve of the relatively dim (V-mag = 11.44) transiting hot Jupiter HAT-P-32b as observed with a single 6" MicroObservatory telescope. (Fowler et al., in prep.)*

All three of these projects are members of NASA UoL's Authentic Ground-based Observing and Research Experiences subgroup, a mix of education and scientific professionals who have experience both in ground-based telescope citizen science projects and professional scientific research, including transiting exoplanet science.

NASA's UoL will also explore partnering with other groups such as the Citizen CATE Experiment, a network of 72 identical 80 mm volunteer operated imaging telescopes distributed across the continental United States, and the Night Sky Network.

Given the expertise of NASA's UoL in working with the public on projects that bring them into data participation and learning and existing projects that would leverage efforts instead of creating new ones, the NASA's UoL team has the experience already in place to make ETS a success.

## Acknowledgements

Part of the research was carried out at the Jet Propulsion Laboratory, California Institute of Technology, under contract with the National Aeronautics and Space Administration. © 2019. All rights reserved. ● NASA's Universe of Learning is supported under award number NNX16AC65A to the Space Telescope Science Institute, working in partnership with Caltech/IPAC, Jet Propulsion Laboratory, Smithsonian Astrophysical Observatory, and Sonoma State University.